\documentclass[lettersize,journal]{IEEEtran}
\usepackage{amsmath,amsfonts}
\usepackage{algorithmic}
\usepackage{algorithm}
\usepackage{array}
\usepackage[caption=false,font=normalsize,labelfont=sf,textfont=sf]{subfig}
\usepackage{textcomp}
\usepackage{stfloats}
\usepackage{url}
\usepackage{verbatim}
\usepackage{graphicx}
\usepackage{cite}
\usepackage{hyperref}
\usepackage{booktabs}
\usepackage{multirow}

\begin{document}

\title{QAMA: Scalable Quantum Annealing Multi‑Head Attention Operator for Deep Learning}

\author{
Peng Du,
Jinjing Shi,~\IEEEmembership{Senior Member IEEE,}
Wenxuan Wang,~\IEEEmembership{Student Member IEEE,}
\\Yin Ma,
Kai Wen,
and Xuelong Li,~\IEEEmembership{Fellow IEEE}
\thanks{This work has been submitted to the IEEE for possible publication. Copyright may be transferred without notice, after which this version may no longer be accessible.}
\thanks{(Corresponding author: Jinjing Shi.)}
\thanks{Peng Du and Jinjing Shi are with the School of Electronic Information, Central South University, China (e-mail: du\_peng@csu.edu.cn, shijinjing@csu.edu.cn, 234701002@csu.edu.cn).}
\thanks{Wenxuan Wang is with the School of Computer Science and Engineering, Central South University, China (e-mail: 234701002@csu.edu.cn).}
\thanks{Yin Ma and Kai Wen are with the Beijing QBoson Quantum Technology Co. Ltd., China (e-mail: may@boseq.com, wenk@boseq.com).}
\thanks{Xuelong Li is with the Institute of Artificial Intelligence (TeleAI) of China Telecom (e-mail: xuelong\_li@ieee.org).}
}

\markboth{Journal of \LaTeX\ Class Files,~Vol.~14, No.~8, August~2021}%
{Shell \MakeLowercase{\textit{et al.}}: A Sample Article Using IEEEtran.cls for IEEE Journals}


\maketitle

\begin{abstract}
Attention mechanisms underpin modern deep learning, while the quadratic time and space complexity limit scalability for long sequences. 
To address this, Quantum Annealing Multi‑Head Attention (QAMA) is proposed, a novel drop-in operator that reformulates attention as an energy-based Hamiltonian optimization problem. 
In this framework, token interactions are encoded into binary quadratic terms, and quantum annealing is employed to search for low-energy configurations that correspond to effective attention patterns. 
Unlike classical sparse or approximate attention methods that rely on hand-crafted heuristics, QAMA allows sparsity structures to emerge naturally from the optimization process.
Theoretically, computational complexity is analysed through single-spin flip dynamics, providing time to solution runtime bounds that depend on the spectral properties of the annealing Hamiltonian. 
Empirically, evaluation on both natural language and vision benchmarks shows that, across tasks, accuracy deviates by at most 2.7 points from standard multi‑head attention, while requiring only linear qubits in sequence length.
Visualizations further reveal that the Hamiltonian penalty terms induce meaningful and interpretable sparsity across heads. 
Finally, deployment on a coherent Ising machine validates the feasibility of running QAMA on real quantum hardware, showing tangible inference-time reductions compared with classical implementations. 
These results highlight QAMA as a pioneering and scalable step toward integrating quantum optimization devices into deep neural architectures, providing a seamlessly integrable and hardware-compatible alternative to conventional attention mechanisms. 
Our implementation is available at \url{https://github.com/Dxee-e/QAMA}.
\end{abstract}


\section{Introduction}
\IEEEPARstart{A}{ttention} mechanism serves as the fundamental basis for models in the contemporary large model paradigm, markedly propelling the advance of generative general artificial intelligence~\cite{introduction-attention-is-all-your-need, introduction-deepseek-r1}.
Specifically, Transformer architectures employ multi-head attention to effectively model intricate contextual dependencies, yielding state-of-the-art results in numerous downstream applications \cite{introduction-transformer-usage1, introdcution-transformer-usage2}.

Despite its effectiveness, classical attention mechanisms face critical scalability challenges.
The quadratic time and space complexity of pairwise similarity computations grows prohibitively with sequence length, creating substantial bottlenecks in both training and inference~\cite{introduction-classical-limit}. 
Numerous strategies have been proposed to mitigate this cost. 
Engineering-level optimizations such as FlashAttention~\cite{FlashAttention} reduce memory overhead, while structural modifications such as Linformer~\cite{Linformer}, Longformer ~\cite{Longformer}, BigBird~\cite{BigBird}, and Performer~\cite{Performer} approximate or sparsify the attention matrix to achieve linear or sub-quadratic scaling.
However, these classical approaches rely on heuristics including fixed windows, hashing, or kernel approximations, and inevitably trade off optimality for efficiency.
As a result, they struggle to provide globally optimal attention patterns and often require hand-designed assumptions about sparsity structure.

From an optimization perspective, attention can be interpreted as a large-scale combinatorial selection problem~\cite{attention_to_combination_optimization}.
Among the vast number of potential token interactions, identifying which ones should be emphasized by each attention head remains a significant challenge.
In particular, structured sparsity is implemented by casting the selection process into a binary quadratic optimization (QUBO) problem, where binary variables encode whether a given token–head pair is selected.
Solving such QUBO problems is NP-hard on classical computers, forcing conventional methods to adopt approximate algorithms or restrictive designs~\cite{introduction-QUBO-NP-problem}. 
This gap motivates exploring alternative computing paradigms that can directly address the optimization nature of attention.

Quantum annealing provides a compelling opportunity in this regard.
By representing the problem as a Hamiltonian, its optimal configuration corresponds to the minimum energy state of the system. 
Quantum annealers, such as coherent Ising machines (CIMs), are specifically designed to solve QUBO problems by evolving toward low-energy states via quantum tunneling~\cite{introduction-CIM-QUBO}. 
This paradigm is advantageous for two primary reasons. 
Firstly, it offers the ability to escape local minima, resulting in more globally optimal attention patterns. 
Secondly, it features a fundamentally different non–Von Neumann architecture that has shown practical runtime improvements on real devices.

In this paper, we introduce Quantum Annealing Multi-Head Attention (QAMA), a novel operator that reformulates multi-head attention as a Hamiltonian optimization challenge and utilizes quantum annealing for its solution.
Unlike classical sparse attention methods that impose heuristic structures, QAMA allows sparsity to emerge naturally from the energy minimization process. 
Our contributions are as follows:

\begin{itemize}
\item Hamiltonian formulation of attention. 
Multi-head attention is cast into a QUBO form, where binary variables represent token–head selection, enabling optimization via quantum annealing.

\item Seamless integration into deep learning. 
QAMA is designed as a drop-in operator compatible with classical models, supporting gradient propagation through implicit differentiation of the energy function.

\item Empirical validation. 
Across NLP and vision benchmarks, QAMA achieves accuracy within 2.7 points of classical attention while delivering significant speedups. 
On real CIM hardware, QAMA reduces inference time from hundreds of milliseconds to microseconds without accuracy loss, demonstrating the practical feasibility of quantum-enhanced attention.
\end{itemize}

The organization of this paper is as follows:
Section \ref{sec:preliminaries} reviews the foundational theory of Quantum Annealing. Section \ref{sec:methodology} describes the QAMA theory and derivations. Section \ref{sec:experiments} presents the experimental configurations and analysis of results. Section \ref{sec:conclusion} concludes the work.

\IEEEpubidadjcol

\section{Preliminaries}\label{sec:preliminaries}

Quantum Machine Learning (QML) is a highly promising paradigm that merges quantum computing with machine learning, demonstrating significant contributions across various domains. 
These include graph theory~\cite{work-PHL, work-QGHNN, work-PEQGNN}, attention mechanisms~\cite{work-QKSAN, work-QSAN}, software testing~\cite{work-QuanTest}, natural language processing~\cite{work-PQIDNNNLP, work-TE2EQIDNNTC, work-QLMEEQA}, classification~\cite{work-IQACLIC, work-HAQJSK, work-HQCCNN, work-few-image-cls}, and finance~\cite{work-QOPLER}.
Quantum annealing (QA) is a rapidly advancing technique for complex optimization, demonstrating versatility by successfully solving industrial problems like the combinatorial optimal power flow (OPF)~\cite{QA-OPF} while also offering foundational hardware insights into multi-qubit coherence~\cite{QA-Dwave}. 
QA computational models have been leveraged in a wide array of domains to achieve better design and optimization.
Notable examples span from Polar Codes~\cite{work-qa-polar-codes} and microgrids~\cite{work-qa-microgrid, work-qa-OEMM, work-qa-grid-para-sim} to path planning~\cite{work-qa-path}.
Extending its utility to machine learning, QA has enabled the creation of novel structures like the quantum annealing hard attention network~\cite{QAHAN}, which utilizes the quantum tunneling effect for faster convergence.

To utilize quantum annealing, the optimization problem need to be converted into finding the ground state of the Ising Hamiltonian. 
This target Hamiltonian is commonly given by Equation~\eqref{eq:qa_ising}.
\begin{align}\label{eq:qa_ising}
\mathcal{H}_{problem}=-\sum_{i<j} \mathcal{J}_{ij} \sigma_i \sigma_j - \sum_{i} \hbar_i \sigma_i
\end{align}
where $\sigma_i \in \{-1,+1\}$ are the spin variables, $\mathcal{J}_{ij}$ denotes the spin-spin coupling strength, and $\hbar_i$ is the external magnetic field acting on spin $i$.

The Quantum Annealer (QA) operates based on the principle of Adiabatic Quantum Computation (AQC)~\cite{qa-aqc, qa-qac-ec}, solving optimization problems by evolving a time-dependent Hamiltonian $\mathcal{H}_t$ in Equation~\eqref{eq:hami-time}:
\begin{align}\label{eq:hami-time}
\mathcal{H}_t=\alpha_t\mathcal{H}_{initial}+\beta_t\mathcal{H}_{problem}
\end{align}
here, $\mathcal{H}_{initial}$ is the initial Hamiltonian and $\mathcal{H}_{problem}$ is the problem Hamiltonian. 
Throughout the process, the coefficient $\alpha_t$ adiabatically decreases from a large value to zero, while $\beta_t$ increases from zero to its maximum value. 
The Adiabatic Theorem mandates that this evolution must be sufficiently slow, thereby guaranteeing the system stays within the instantaneous ground state of $\mathcal{H}_t$.

In contrast, CIM is an optical system based on DOPOs, whose operation relies on classical nonlinear dynamics~\cite{cim-1, cim-2, cim-3, CIM-OPO}. 
The CIM guides the system state, represented by continuous soft spins, to dynamically converge to a low-energy configuration of $\mathcal{H}_{problem}$ by gradually increasing the laser gain. 
Although quantum-inspired, the CIM achieves an annealing-like optimization process through a mechanism known as Geometric Landscape Annealing to find the optimal solution of the target Hamiltonian.

\section{Methodology}\label{sec:methodology}

\subsection{Hamiltonian of Energy-Quantized Attention Mechanism}

The energy-based attention mechanism is designed to facilitate applications on quantum annealing computers by reformulating the multi-head attention mechanism into a Hamiltonian form.
In this architecture, the attention mechanism is modeled as a three-body system defined by the interaction among the query $\mathcal{Q}$, key $\mathcal{K}$, and value $\mathcal{V}$ feature vectors that constitute the system's state.
Initially, the system is in an excited state, implying a strong focus on each feature.
Through an adiabatic annealing process, the system evolves to its energy ground state, yielding a stable eigenstate that corresponds to the optimal attention configuration.
QAMA models the $\mathcal{Q}$, $\mathcal{K}$, and $\mathcal{V}$ as Hamiltonians, where eigenvalues of the Hamiltonian ground state produced by the model represent the attention mechanism's optimal field of view.
Specifically, the optimal feature vector corresponds to the maximum effective token selection of the attention mechanism when operating in a multi-head constraints.

Inspired by the aforementioned motivation, the Hamiltonian expression for QAMA is constructed.
Let the tensor $\mathcal{Q} \in \mathbb{R}^{B \times H \times N \times D}$,
the tensor $\mathcal{K} \in \mathbb{R}^{B \times H \times N \times D}$, 
and the tensor $\mathcal{V} \in \mathbb{R}^{B \times H \times N \times D}$.
The key variables are defined as follows: $B$ for batch size, $H$ for the number of heads, $N$ for the sequence length, and $D$ for the dimension of each feature vector.
For tractability, assume the condition that the dimensions of the Query $\mathcal{Q}$ are equivalent to those of the Key $\mathcal{K}$ and Value $\mathcal{V}$. 
Such a constraint holds true for most practical applications and substantially reduces the complexity of the theoretical formulation.
Then, the problem is formally restated within the Hamiltonian formalism.
The Hamiltonian of the QAMA operator is defined as follows in Equation~\eqref{eq:hamiltonian}.
\begin{align}\label{eq:hamiltonian}
\left\{\begin{aligned}
\mathcal{H} & = -\mathcal{H}_{\alpha} - \rho \mathcal{H}_{\beta} + \lambda \mathcal{H}_{\gamma} \\
\mathcal{H}_{\alpha} & = \sum_{t=1}^{H} \sum_{i=1}^{N} \sum_{j=1}^{N} \mathcal{J}_{t,i,j} \cdot s_{t,i} \cdot s_{t,j}\, , \, i < j \\
\mathcal{H}_{\beta} &= \sum_{t=1}^{H} \sum_{i=1}^{N} \hbar_{t,i} \cdot s_{t,i} \\
\mathcal{H}_{\gamma} &= \sum_{i=1}^{N} \sum_{t_1=1}^{H} \sum_{t_2=1}^{H} s_{t_1,i} \cdot s_{t_2,i}\, , \, t_1 < t_2 \\
\end{aligned}\right.
,\end{align}
where $\mathcal{H}$ represents the Hamiltonian of multi-head attention system, $\mathcal{H}_{\alpha}$ is a subterm representing interactive energy, $\mathcal{H}_{\beta}$ is a subterm for linear importance, and $\mathcal{H}_{\gamma}$ is a penalty term designed for sparse heads within the multi-head mechanism. 
The variable $s \in \{0,1\}^{H \times N}$ is a binary coefficient and $s_{t, i}$ corresponding to the $t$-th head and $i$-th feature. 

For the quadratic term $\mathcal{H}_{\alpha}$, $\mathcal{J}_{t,i,j}$ is designed to uncover the relationships between feature $\mathcal{Q}_{t,i,*}$ and feature $\mathcal{K}_{t,j,*}$, ensuring its symmetric characteristic. 
For the linear term $\hbar_{t,i}$, its purpose is to represent the self-importance of feature $\mathcal{V}_{t,i,*}$.
The calculation formula is shown in Equation~\eqref{eq:Jh}.
For the penalty term $\mathcal{H}_{\gamma}$ is designed to enforce sufficient dissimilarity across the heads, thus mitigating the emergence of redundant feature learning processes. 
Its formulation as an energy optimization objective ensures that head differentiation and feature learning are achieved concurrently within a unified optimization procedure.
\begin{align}\label{eq:Jh}
\left\{\begin{aligned}
\mathcal{J}_{t,i,j}& = \frac{\mathcal{Q} \mathcal{K}^T \mathcal{K} \mathcal{Q}^T}{2D} \, , \,  i \neq j\\
\hbar_{t,i}& = \mathcal{V}_{t,i,*} W_\epsilon
\end{aligned}\right.
\end{align}
where $W_\epsilon \in \mathbb{R}^{D\times 1}$ is a learnable parameter that transforms a vector into an important energy scalar. 

The Hamiltonian defined by Equation~\eqref{eq:hamiltonian} is utilized to optimize the energy system, seeking to determine the system's ground state. 
Upon achieving this ground state, the corresponding lowest energy eigenvalue is yielded.
In order to maintain dimensional consistency between the output features of QAMA and the conventional attention mechanism, which is crucial for model interoperability (plug-and-play), the energy scalar is remapped into the feature space through a specialized distribution mapping transformation.
The output of the entire attention mechanism will be adjusted from the raw energy $\mathcal{E}_{out} \in \mathbb{R}$ to energy component output $\mathcal{E}_{dist} \in \mathbb{R}^{D}$ based on the distribution of $W_\epsilon$, shown in Equation~\eqref{eq:map}. 
The output energy encapsulates both the interaction information between features and their individual importance.
\begin{align}\label{eq:map}
\left\{\begin{aligned}
\mathcal{E}_{out} &= - \mathcal{H}_{\alpha} - \rho \mathcal{H}_{\beta} \\
\mathcal{E}_{dist} &= \mathcal{E}_{out} W_{\epsilon}^{T}
\end{aligned}\right.
.\end{align}

\subsection{Non-differentiable Quantum Annealing Layer Gradient Propagation}

A key challenge in integrating quantum annealing layers into neural networks is their non-differentiability, which prevents the direct application of automatic differentiation as implemented in modern frameworks.
Consequently, a strategy was developed that combines the technique of reparameterization~\cite{reparameterization-trick} with stop-gradient operations~\cite{stop-gradient-op-0, stop-gradient-op-1}.
This approach enables the propagation of gradients through the energy-based attention mechanism module during the forward pass.
The backpropagation process is formally described by the following partial derivatives of the distributed output energy $\mathcal{E}_{dist}$ with respect to the input features $\mathcal{T}$.
The backpropagation is shown in Equation~\eqref{eq:backpropagation}.
\begin{align}\label{eq:backpropagation}
\left\{\begin{aligned}
\frac{\partial \mathcal{E}_{dist}}{\partial \mathcal{T}} &= \frac{\partial \mathcal{E}_{dist}}{\partial \mathcal{E}_{out}} \cdot (-\frac{\partial \mathcal{E}_{out}}{\partial \mathcal{H}_{\alpha}} \cdot \frac{\partial \mathcal{H}_{\alpha}}{\partial \mathcal{T}} - \rho \frac{\partial \mathcal{E}_{out}}{\partial \mathcal{H}_{\beta}} \cdot \frac{\partial \mathcal{H}_{\beta}}{\partial \mathcal{T}}) \\
\frac{\partial \mathcal{H}_{\alpha}}{\partial \mathcal{T}} &= \frac{\partial \mathcal{J}(\mathcal{T})}{\partial \mathcal{T}} \cdot (s' \cdot s)^{\text{detached}} \\
\frac{\partial \mathcal{H}_{\beta}}{\partial \mathcal{T}} &= \frac{\partial \hbar(\mathcal{T})}{\partial \mathcal{T}} \cdot s^{\text{detached}}
\end{aligned}\right.
,\end{align}
where $s^\text{detached}$ denotes gradient truncation. 
The gradient is implicitly solved via the attention energy function, thereby skipping the direct gradient calculation of the non-differentiable layer and enabling gradient reparameterization.

\subsection{Architecture of QAMA}

\begin{figure*}[t!]
    \centering
    \includegraphics[width=0.95\linewidth]{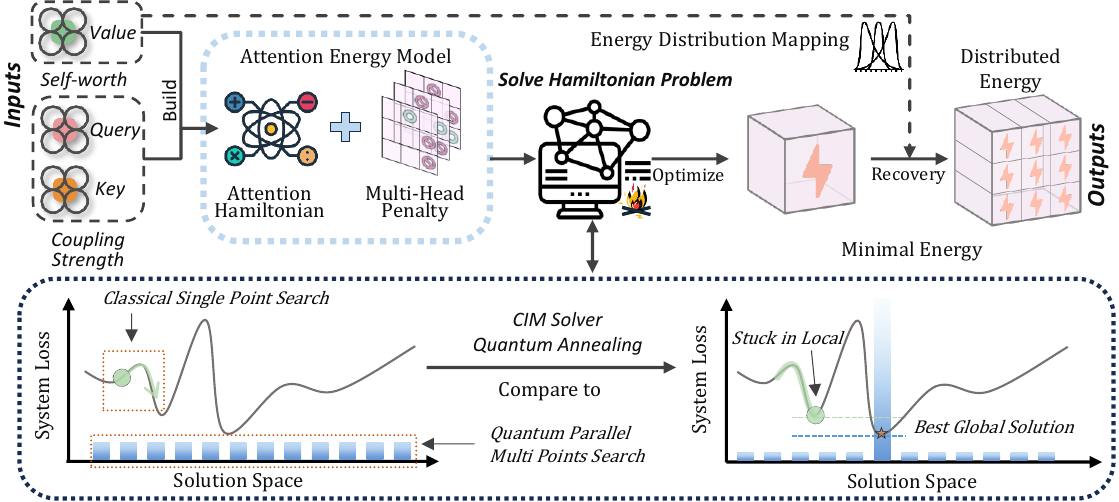}
    \caption{The quantum annealing solves the quantized attention mechanism model to obtain the optimal attention solution.}
    \label{fig:qama-arch}
\end{figure*}

As shown in Figure \ref{fig:qama-arch}, the overall QAMA architecture is presented. 
The core QAMA operator is primarily divided into three functional components:
(1) Quantized attention mechanism problem construction.
This initial stage involves decomposing the energy Hamiltonian of the attention mechanism into three subterms using the previously described method. 
This process effectively transforms the dynamic computation of attention weights into an energy minimization problem.
(2) Quantum annealing solver. 
The QUBO problem is solved using a CIM (Coherent Ising Machine) quantum annealer. 
This fundamentally distinct computational architecture offers a significant reduction in both time and space complexity compared to classical computing methods.
(3) Energy distribution mapping for output features.
In the final stage, the energy scalars are transformed back into their respective energy components through an inverse distribution process, ensuring consistency in the output features.

During the backward propagation phase, gradients are implicitly transmitted through the meticulously constructed energy equation, thereby ensuring the differentiability of the quantum annealing layer.
The comprehensive description of the pseudo-algorithm can be found in Algorithm~\ref{alg:qama}.
QAMA is designed to be implemented as a deep learning operator, allowing for a seamless replacement of classical attention mechanisms. 
While maintaining performance levels consistent with classical approaches, it substantially reduces time and space complexity. 
This architecture realizes a heterogeneous attention mechanism on a quantum annealing computer, leveraging its non-Von Neumann computing architecture.

\begin{algorithm}
\caption{QAMA Module Implement}
\label{alg:qama}
\begin{algorithmic}[1]
\STATE \textbf{Input:} Query tensor $\mathcal{Q}$, Key tensor $\mathcal{K}$, Value tensor $\mathcal{V}$
\STATE \textbf{Output:} Distributed energy $E_{\text{dist}}$
\STATE The quadratic term $H_\alpha$, the linear term $H_\beta$, and the penalty term $H_\gamma$ are calculated.
\STATE Calculate Dynamic Coefficient Scaling $\rho$ and $\lambda$.
\IF{Current in forward propogation}
\STATE Initialize Hamiltonian $H = -H_\alpha - \rho H_\beta + \lambda H_\gamma$.
\STATE The Hamiltonian is transformed into a QUBO model.
\STATE The resultant QUBO model is converted into the Ising Model format supported by the CIM architecture.
\STATE Solve the problem using CIM quantum annealer to find ground state $s_{t,i}$.
\ENDIF
\STATE Using ground state to compute output energy $E_{\text{out}} = -H_\alpha - \rho H_\beta$, without penalty.
\STATE Mapping the energy from a scalar representation $E_{\text{out}}$ into a reconstructed distribution $E_{\text{dist}}$.
\RETURN Distributed energy $E_{\text{dist}}$
\end{algorithmic}
\end{algorithm}

\subsection{Coefficient Proof}
The coefficients $\rho$ and $\lambda$ in the QAMA Hamiltonian are determined by relative magnitudes of the interaction term $\mathcal{H}_{\alpha}$, the linear term $\mathcal{H}_{\beta}$, and the penalty term $\mathcal{H}_{\gamma}$.
Therefore, a reasonable estimation of the maximum values for each Hamiltonian subterm is essential.
Given that the $\mathcal{Q}, \mathcal{K}, \mathcal{V}$ tensors have been normalized, their value ranges can be approximately estimated using the expectation of normal distribution.

Let $X \sim N(0,1)$. We know that $\mathbb{P}(X>0)=0.5$. Under the condition that $X>0$, the expectation is Equation~\eqref{eq:cp}.
\begin{equation}\label{eq:cp}
\mathbb{E}[X \mid X > 0] = \frac{\int_{0}^{\infty} x \cdot \frac{1}{\sqrt{2\pi}} e^{-x^2/2} dx}{\int_{0}^{\infty} \frac{1}{\sqrt{2\pi}} e^{-x^2/2} dx} = \sqrt{\frac{2}{\pi}}
.\end{equation}

For the purpose of calculating the value range of the sub-terms using conditional probability, we omit the self-variable s to ensure the independence of each sub-term. 
This approach allows the problem to be equivalently solved by determining the maximum expected value of the terms.
For the penalty term $\mathcal{H}_{\gamma}$, the resulting maximum expected value is equivalent to the number of possible head combinations by $H$ choosing 2.
For interactive and linear terms, the expected maximum value can be expressed as Equation~\eqref{eq:cp2}.
\begin{equation}\label{eq:cp2}
\mathbb{E}[\mathcal{H}_{*}^{max}]=length(X) \cdot \mathbb{E}[X \mid X > 0] \cdot \mathbb{P}(X>0)
.\end{equation}

Thus, the expected maximum value for each subterm is given by Equation~\eqref{eq:range}.
\begin{align}\label{eq:range}
\left\{\begin{aligned}
\mathbb{E}[\mathcal{H}_{\alpha}^{max}]&=H \cdot N^2 \cdot \sqrt{\frac{1}{2\pi}} \\
\mathbb{E}[\mathcal{H}_{\beta}^{max}]&=H \cdot N \cdot \sqrt{\frac{1}{2\pi}} \\
\mathbb{E}[\mathcal{H}_{\gamma}^{max}]&=\binom{H}{2} \cdot N \\
&=\frac{H(H-1) \cdot N}{2}
\end{aligned}\right.
.\end{align}

Based on these proportional relationships, the dynamic coefficients for the linear term ($\rho$) and the penalty term ($\gamma$) are derived as follows Equation~\eqref{eq:hp}.
\begin{align}\label{eq:hp}
\left\{\begin{aligned}
\rho &= N \cdot \rho_0 \\
\gamma &= \frac{N \cdot \sqrt{\frac{2}{\pi}}}{H - 1} \cdot \gamma_0
\end{aligned}\right.
,\end{align}
where, $\rho_0$ and $\gamma_0$ are static coefficients within the range $[0, 1]$, whose specific values are determined during experimental setup. 
This methodology ensures a proper dynamic balance in the value ranges across the different terms of the Hamiltonian.

\subsection{QAMA Time and Space Complexity Proof}

\textbf{Space Complexity.}
In a CIM, space complexity is determined by the number of physical components, specifically optical parametric oscillators (OPOs), which form the quantum annealing system. 
For QAMA, the architecture necessitates $H \cdot N$ qubits, with each qubit realized by an OPO. 
Furthermore, the couplings are integrated into the system design and do not require additional memory~\cite{complexity-couple}.
Consequently, the space complexity of QAMA is $O(H \cdot N)$, demonstrating a linear scaling with increasing of problem size.
This capability can be attributed to the quantum superposition states of the qubits inherent to Quantum Annealing, facilitating the representation of $N^2$ inter-feature information using only $N$ qubits.

\textbf{Time Complexity.}
The CIM is realized through an optical-feedback mechanism that inherently includes dissipation, noise, and a measurement-feedback channel~\cite{CIM-OPO}. 
The dynamical model employed here utilizes a classical single-spin flip transition probability, typical of Metropolis or Glauber schemes~\cite{MG-transition-probability}.

Initially, the QUBO-based Hamiltonian is transformed into the Ising model representation. 
The conversion to the form of Equation~\eqref{eq:ising} is achieved by applying the change of basis defined as $x=\frac{1+\sigma}{2}$.
\begin{align}\label{eq:ising}
\left\{\begin{aligned}
\mathcal{H} & = -\mathcal{H}_{\alpha} - \rho \mathcal{H}_{\beta} + \lambda \mathcal{H}_{\gamma} \\
\mathcal{H}_{\alpha} & = \sum_{t=1}^{H} \sum_{i=1}^{N} \sum_{j=1}^{N} \mathcal{J}_{t,i,j} \cdot \frac{(1+\sigma_{t,i})(1+\sigma_{t,j})}{4} \, , \, i < j \\
\mathcal{H}_{\beta} &= \sum_{t=1}^{N} \sum_{i=1}^{M} \hbar_{t,i} \cdot \frac{1+\sigma_{t,i}}{2} \\
\mathcal{H}_{\gamma} &= \sum_{i=1}^{N} \sum_{t_1=1}^{H} \sum_{t_2=1}^{H} \frac{(1+\sigma_{t,i})(1+\sigma_{t,j})}{4}\, , \, t_1 < t_2 \\
\end{aligned}\right.
,\end{align}

For the spin variable $\sigma_{p,k} \in \{-1,+1\}$ which represente $p$-th head and $k$-th feature, the energy difference caused by flipping it can be expressed as Equation~\eqref{eq:flip-1}.
\begin{align}\label{eq:flip-1}
\left\{\begin{aligned}
\Delta \mathcal{H}^{-}_{p,k} &= - \sum_{i=1,i\neq k}^{N} \mathcal{J}_{p,k,i} \cdot \frac{\sigma_{p,i}+1}{2} - \rho \cdot \hbar_{p,k} \\&\quad + \gamma \cdot \sum_{t=1,t\neq p}^{H} \frac{\sigma_{t,k}+1}{2} \\
\Delta \mathcal{H}^{+}_{p,k} &= \sum_{i=1,i\neq k}^{N} \mathcal{J}_{p,k,i} \cdot \frac{\sigma_{p,i}+1}{2} + \rho \cdot \hbar_{p,k} \\&\quad - \gamma \cdot \sum_{t=1,t\neq p}^{H} \frac{\sigma_{t,k}+1}{2} = -\Delta \mathcal{H}^{-}_{p,k} \\
\end{aligned}\right.
\end{align}
where $\Delta H^{-}_{p,k}$ denotes the flipping energy difference when the spin variable $\sigma_{p,k}$ changes from -1 to +1, and $\Delta H^{+}_{p,k}$ denotes the flipping energy difference when the spin variable $\sigma_{p,k}$ changes from +1 to -1.
which can be compactly rewritten as
\begin{align}
\Delta \mathcal{H} = \sigma_{p,k}' \cdot \Delta \mathcal{H}^{-}_{p,k}
\end{align}
where $\sigma_{p,k}'$ denotes the state after flipping, and $\delta \mathcal{H}$ is a general expression of energy difference by flipping 1 bit.

The explicit bounds on the annealing time $T$ using the single-spin flip energy differences defined in Equation~\eqref{eq:flip-1} is derived.
To transition to the global ground state, a certain number of spins must be reversed along a flip path starting from the initial state.
The energy barrier along this path determines its dominant exponential cost. 
For any given path $\pi=(v_1, \dots, v_m)$, where $v$ represents the spin index, the path energy barrier is defined as:
\begin{align}
B(\pi)=\max_{1\le r \le m} \max\{0,\Delta \mathcal{H}(v_r \text{ at step }r)\}
\end{align}

Therefore, the minimum possible energy barrier needed to change the configuration to the ground state is given by:
\begin{align}
B_{min}=\min_{\text{possible path } \pi} B(\pi)
\end{align}
The quantity $B_{min}$ defines the minimum energy barrier required for a trajectory to transition to the ground state.
When a path involves several steps demanding a positive energy increase, the summation of these increments provides a conservative upper bound for the overall total energy barrier.
\begin{align}
B_{min} = \min_{\pi} \sum_{r:\Delta>0} \Delta \mathcal{H}(v_r)
\end{align}

Assuming that the effective inverse temperature of the system increases to $\beta$ during a certain period of the annealing process, the transition probability from the current state to a candidate states can be written as:
\begin{align}\label{eq:time_p}
P \approx \Theta(e^{-\beta B_{min}})
\end{align}
Based on the empirical relationship~\cite{time2solution}, the computation time is determined to be:
\begin{align}
T_{sol} = T_{ann} \cdot \left\lceil \frac{\ln{0.01}}{\ln(1-P)} \right\rceil
\end{align}
where $T_{sol}$ is total expected solution time, which is related to the single-run annealing time $T_{ann}$ and the single-run success probability $P$.
By utilizing the approximation $\ln(1-P)\approx -P$ for small success probabilities $P\ll 1$,$T_{sol}$ is simplified to: 
\begin{align}
T_{sol}\approx T_{ann} \cdot \frac{\ln{0.01}}{-P} = \Theta(T_{ann} P^{-1})
\end{align}
according to the~\ref{eq:time_p}, The $T_{sol}$ will finnaly be the given formation:
\begin{align}
T_{sol} = \Theta(T_{ann} e^{\beta B_{min}})
\end{align}

Assume there exists a path $\pi^*$ whose path energy barrier is $B_{min}$. Then, any annealing algorithm based on single-spin flips requires a running time of at least:
\begin{align}
T_{sol} = \Omega(T_{ann} e^{\beta B_{min}})
\end{align}

Assume there exists a path $\pi$ whose cumulative positive energy sum does not exceed $B_u$. Then, a conservative achievable upper bound for the total solution time is:
\begin{align}
T_{sol} = O(T_{ann}C e^{B_u})
\end{align}
where $C$ represents the factor containing the polynomial path count, which can be expressed as poly($n$). 
The energy barrier $B_u$ is bounded by $B_u\le m\cdot \max\{0,\Delta \mathcal{H}\}$, where $m$ is the number of spins that need to be flipped to reach the ground state from an arbitrary state, and $m<H\cdot N$.

In summary, the time complexity and its bounds for QAMA were derived using the Time-to-Solution (TTS) method.
It is evident from this analysis that the time complexity is independent of the scale of the problem.
However, the complexity is directly related to the intrinsic nature of the problem. 
Since different problems correspond to distinct ground-state paths and energy gaps, the required running time is inherently problem-dependent.
Although the resultant time complexity is exponential, this calculation reflects the number of running cycles of the underlying Coherent Ising Machine (CIM) hardware, rather than the number of iterations used in classical algorithm analysis.
Consequently, this time complexity is not comparable to conventional calculation metrics used in the domain of classical computing.
The objective of this proof is to demonstrate that the time complexity is independent to scale and is only defined by the specific form of the problem Hamiltonian $\mathcal{H}_{problem}$.
To provide stronger evidence, we experimentally measured the specific running time, with the CIM solution time being experimentally verified in Section~\ref{sec:cim_time}.

\section{Experiments}\label{sec:experiments}
In this section, a set of experiments designed to address the following key questions concerning the QAMA operator and its performance on CIM and simulation device:

\begin{itemize}
\item How well does the QAMA operator generalize? Can it effectively adapt to various downstream tasks while demonstrating performance comparable to classical attention mechanisms?

\item What is the practical performance of QAMA when deployed on a real CIM quantum annealing hardware?

\item What specific advantages does QAMA offer over other attention mechanisms that are also optimized for spatiotemporal complexity?
\end{itemize}

The model hyperparameters for QAMA were determined through the guiding principle that the ratio of interaction terms to linear terms should approximate 1.3, and the ratio of interaction terms to penalty terms should be approximately 2.
This process yielded the hyperparameters $\rho_0=0.16$ and $\gamma_0=0.8$.
The training of QAMA is performed using a simulated annealing algorithm.
The specific experimental configurations for addressing each research question are detailed in different question.

\subsection{QAMA quantum operator performance consistency and generalization}

\begin{table}[htbp]
\centering
\caption{Fundamental Configurations for the core task}
\label{tab:config}
\resizebox{\columnwidth}{!}{
\begin{tabular}{cccccc}
\hline
Model &
  \multicolumn{3}{c}{SimpleViT} &
  \multicolumn{2}{c}{\begin{tabular}[c]{@{}c@{}}Text\\ Transformer\end{tabular}} \\ \hline
Dataet &
  MNIST &
  CIFAR10 &
  \begin{tabular}[c]{@{}c@{}}Fashion\\ MNIST\end{tabular} &
  IMDB &
  SST2 \\
\begin{tabular}[c]{@{}c@{}}Shape\\ Max Length\end{tabular} &
  1$\times$28$\times$28 &
  3$\times$32$\times$32 &
  1$\times$28$\times$28 &
  512 &
  512 \\
Patch Size & 4$\times$4 & 4$\times$4 & 4$\times$4 & -         & -         \\
Qubits     & 392        & 512        & 392        & 512       & 512       \\
Heads      & \multicolumn{3}{c}{8}                & \multicolumn{2}{c}{1} \\
\begin{tabular}[c]{@{}c@{}}Learning \\ Rate\end{tabular} &
  \multicolumn{5}{c}{0.001} \\
Batch Size & \multicolumn{5}{c}{64; 128}                                  \\
Optimizer  & \multicolumn{5}{c}{Adam Optimizer}                           \\
Epoches    & \multicolumn{5}{c}{50}                                       \\ \hline
\end{tabular}
}
\end{table}

\begin{table}[htbp]
\centering
\caption{Extended Experimental Configurations for generalization and performance consistency}
\label{tab:config_extend}
\resizebox{\columnwidth}{!}{
\begin{tabular}{ccccc}
\hline
\multicolumn{2}{c}{Variables} &
  MNIST &
  CIFAR10 &
  FashionMNIST \\ \hline
\multirow{3}{*}{\begin{tabular}[c]{@{}c@{}}Heads\\ Numbers\end{tabular}} &
  Patch Size &
  \multicolumn{3}{c}{4$\times$4} \\
 &
  Heads &
  \multicolumn{3}{c}{1 / 2 / 4 / 8} \\
 &
  Qubits &
  \begin{tabular}[c]{@{}c@{}}49 / 98 /\\ 196 / 392\end{tabular} &
  \begin{tabular}[c]{@{}c@{}}64 / 128 /\\ 256 / 512\end{tabular} &
  \begin{tabular}[c]{@{}c@{}}49 / 98 /\\ 196 / 392\end{tabular} \\ \hline
\multirow{4}{*}{\begin{tabular}[c]{@{}c@{}}Token\\ Length\end{tabular}} &
  Patch Size &
  \multicolumn{3}{c}{2$\times$2} \\
 &
  Heads &
  \multicolumn{3}{c}{2} \\
 &
  Average Pool &
  \multicolumn{3}{c}{1$\times$1 / 2$\times$2 / 2$\times$7 / 7$\times$7 / 14$\times$14} \\
 &
  Length &
  \begin{tabular}[c]{@{}c@{}}196 / 49 /\\ 14 / 4 / 1\end{tabular} &
  \begin{tabular}[c]{@{}c@{}}256 / 64 /\\ 16 / 4 / 1\end{tabular} &
  \begin{tabular}[c]{@{}c@{}}196 / 49 /\\ 14 / 4 / 1\end{tabular} \\ \hline
\end{tabular}
}
\end{table}

For downstream CV tasks, we train the SimpleViT model~\cite{SimpleViT} on the MNIST~\cite{MNIST}, FashionMNIST~\cite{FashionMNIST}, and CIFAR10~\cite{CIFAR} datasets.
For downstream NLP tasks, the TextTransformer model is trained on the IMDB~\cite{IMDB} and SST2~\cite{SST2} datasets.
The extra fundamental parameters used for this task are listed in Table~\ref{tab:config}.
We separately evaluate the classification accuracy of both the original attention mechanism and the QAMA operator-replaced version. 
Additionally, the mask and energy states of QAMA are extracted and visualized for in-depth analysis.
In order to conduct a more thorough examination of generalization and performance alignment with classical baseline, we designed a two-pronged experimental approach.
Initially, experiments conducted by adjusting the heads number to assess the accuracy changes and differences exhibited by QAMA relative to the baseline. 
The configurations utilized for this initial study are presented in Table~\ref{tab:config_extend}.
Subsequently, we modified the input sequence length fed to the attention mechanism via Average Pooling. 
This step serves to validate the shifts and disparities in performance between QAMA and the baseline when processing inputs of various lengths. 
The specific experimental settings are outlined in Table~\ref{tab:config_extend}.

\begin{table}[htbp]
\centering
\caption{Performance consistency comparison table of QAMA.}
\label{tab:QAMA}
\resizebox{\columnwidth}{!}{
\begin{tabular}{@{}cclccc@{}}
\toprule
\multirow{2}{*}{Model} &
  \multirow{2}{*}{Dataset} &
   &
  \multicolumn{2}{c}{Accuracy} &
  \multirow{2}{*}{Delta} \\ \cmidrule(lr){4-5}
 &
   &
   &
  Base &
  QAMA &
   \\ \midrule
\multirow{6}{*}{SimpleViT} &
  \multirow{2}{*}{MNIST} &
  * &
  96.9\% &
  95.3\% &
  -1.6\% \\
 &
   &
  \# &
  97.4\% &
  95.9\% &
  -1.5\% \\ \cmidrule(l){2-6} 
 &
  \multirow{2}{*}{CIFAR10} &
  * &
  61.4\% &
  59.3\% &
  -2.1\% \\
 &
   &
  \# &
  62.0\% &
  59.3\% &
  -2.7\% \\ \cmidrule(l){2-6} 
 &
  \multirow{2}{*}{\begin{tabular}[c]{@{}c@{}}Fashion\\ Mnist\end{tabular}} &
  * &
  86.7\% &
  86.1\% &
  -0.6\% \\
 &
   &
  \# &
  87.7\% &
  86.7\% &
  -1.0\% \\ \midrule
\multirow{4}{*}{\begin{tabular}[c]{@{}c@{}}Text\\ Transformer\end{tabular}} & \multirow{2}{*}{IMDB} & * & 92.6\% & 92.6\% & 0.0\% \\
 &
   &
  \# &
  92.6\% &
  92.8\% &
  0.2\% \\ \cmidrule(l){2-6} 
 &
  \multirow{2}{*}{SST2} &
  * &
  89.2\% &
  87.5\% &
  -1.7\% \\
 &
   &
  \# &
  90.4\% &
  89.6\% &
  -0.8\% \\ \bottomrule
\end{tabular}
}
\end{table}

An analysis of comparative results shown in Table~\ref{tab:QAMA} reveals that QAMA operator achieves accuracy comparable to its classical counterpart and demonstrates the potential to surpass it in certain scenarios.
The symbol * indicates the last epoch, and the symbol \# indicates the best epoch.
This establishes QAMA as a viable and promising alternative to standard attention mechanisms.

\begin{figure}[htbp]
    \centering
    \includegraphics[width=0.85\linewidth]{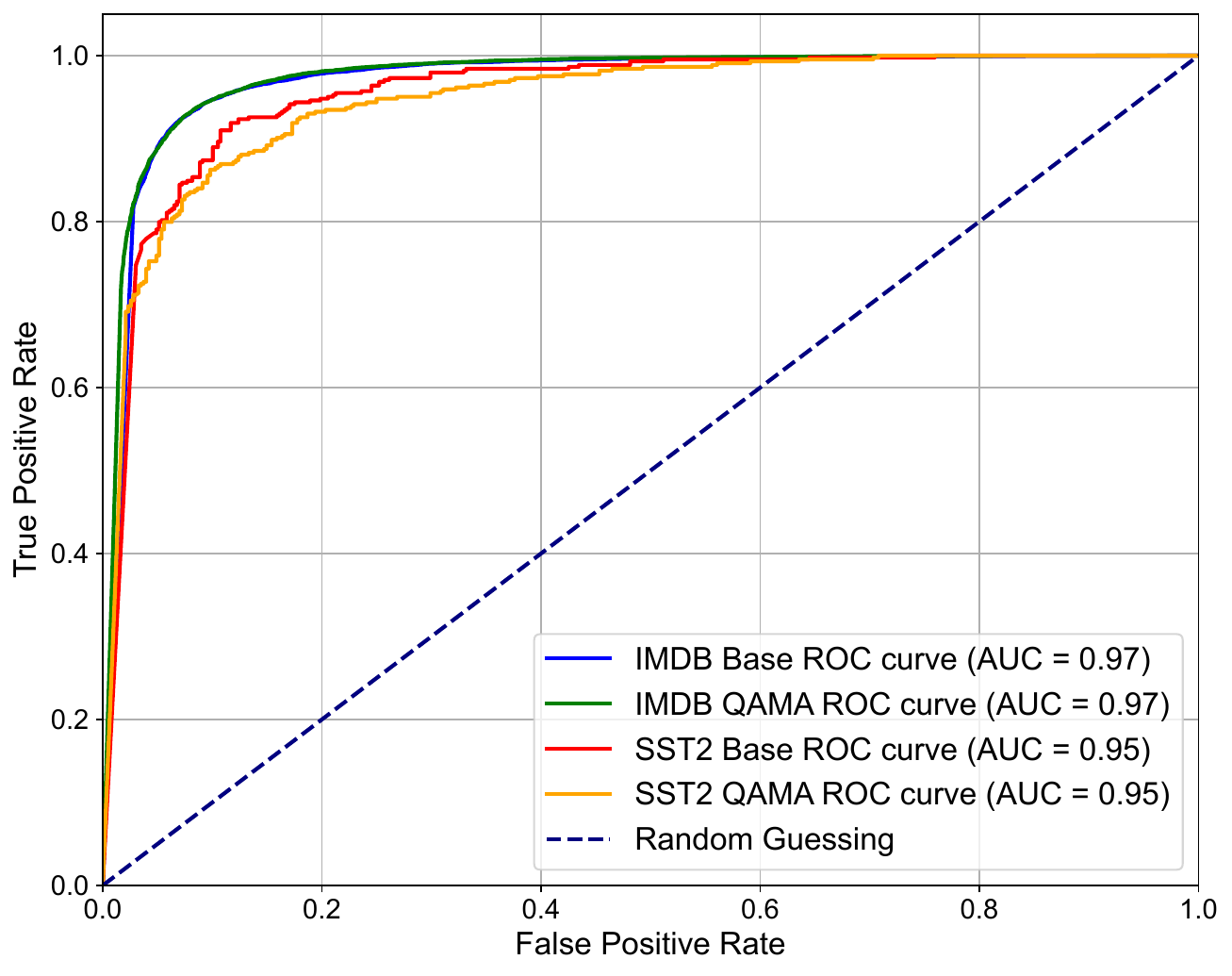}
    \caption{ROC and AUC curve for binary classification NLP tasks of IMDB and SST-2.}
    \label{fig:roc}
\end{figure}

Futuremore, the operator exhibited better accuracy in the NLP domain.
On IMDB dataset, TextTransformer with QAMA-replaced not only matched but exceeded the baseline's best-epoch accuracy by +0.2\%.
This result successfully demonstrates QAMA's capability to offer a distinct performance advantage. 
Furthermore, on the SST2 dataset, the operator remained highly competitive, with its best-epoch accuracy showing only a marginal difference of -0.8\% from the classical model, confirming its ability on performance with the baseline.
The ROC and AUC curves of Figure~\ref{fig:roc} demonstrate the excellence of its effect and the consistency after QAMA replacement.

In the CV domain, the QAMA operator showed strong functional adaptability within the SimpleViT architecture. 
While the results indicate a performance gap when compared to the classical baseline—ranging from -1.0\% on FashionMNIST to -2.7\% on CIFAR10, the operator consistently maintained a similar accuracy.
This suggests that while classical attention currently shows an advantage in these specific vision tasks, QAMA serves as a robust proof-of-concept whose performance can be further optimized.

To further analyze the effectiveness and interpretability of the QAMA mechanism, we conducted detailed visualization experiments on a randomly selected image from the CIFAR-10 test set.

\begin{figure}[htbp]
    \centering
    \includegraphics[width=\linewidth]{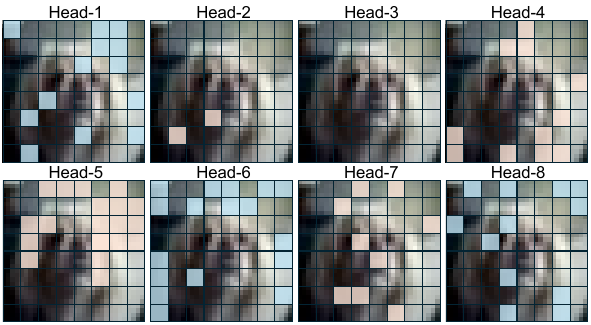}
    \caption{Qubit-based head mask visualization in multi-head mechanism.}
    \label{fig:mask}
\end{figure}

The sparse distributions of different heads are shown in Figure~\ref{fig:mask}, which displays the qubit-based head mask activation. 
Each head selectively attends to different semantic parts of the input image. 
For example, Head-4 is focused on the dog's hair region, while Head-7 is concentrated around the eyes, nose, and mouth. 
This indicates the sparsity and disentanglement in attention across heads, which arise from the penalty term in the QAMA Hamiltonian. 

\begin{figure}[htbp]
    \centering
    \includegraphics[width=\linewidth]{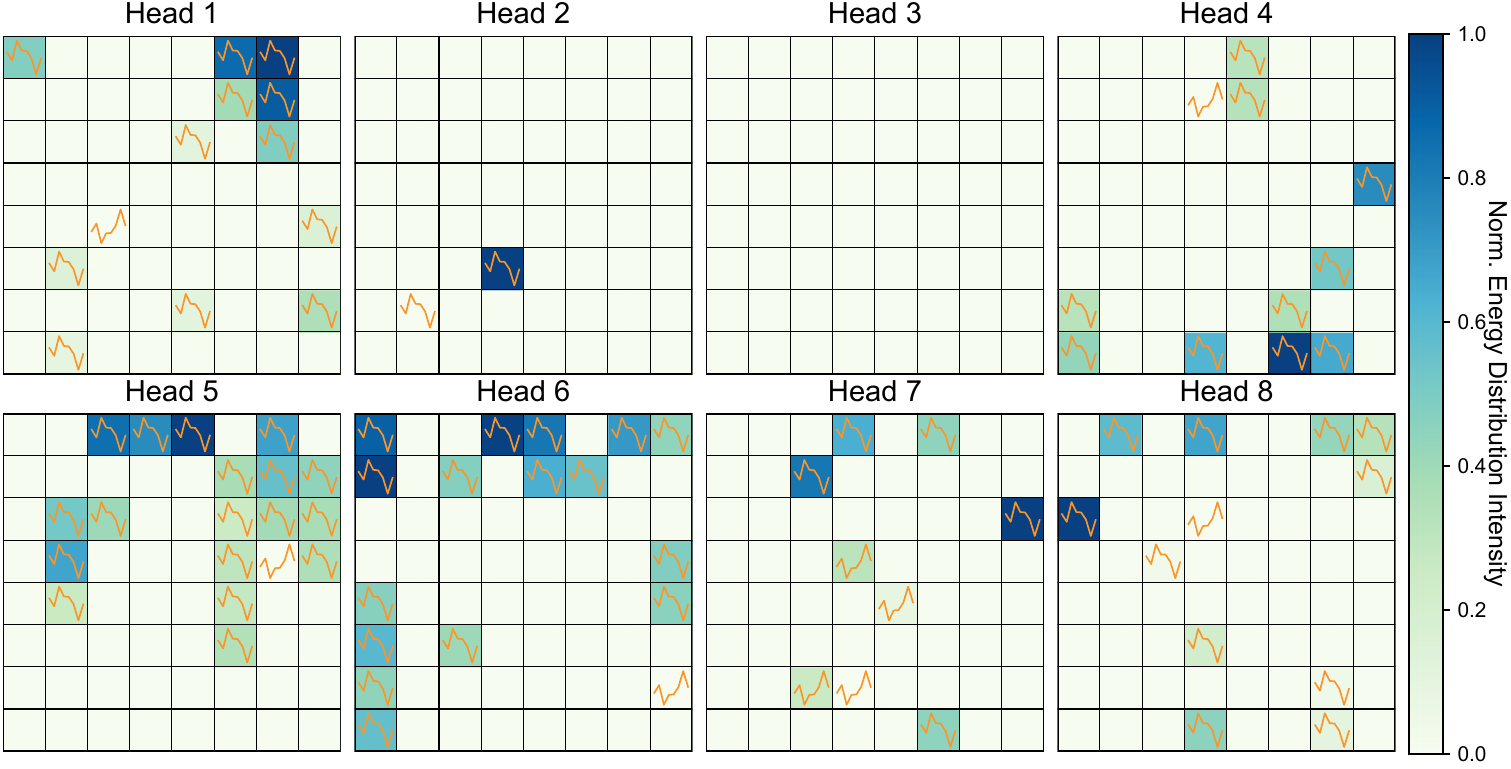}
    \caption{Energy-based attention mechanism output value energy average heat map and feature distribution}
    \label{fig:energy}
\end{figure}

The energy-based feature mapping is presented by the heat map Figure~\ref{fig:energy}.
By projecting the scalar energy into the image feature space, the system's total pure energy is -143.88. 
Following projection, the data adopts a distribution characterized by a mean of 0.00 and a standard deviation of 0.51, thus successfully matching the structure of the initial normalized input tensor.
This demonstrates QAMA's ability to retain and propagate informative representations between layers, even under the constraints of quantum annealing, ensuring continuity in attention-based learning.

\begin{figure}[htbp]
    \centering
    \includegraphics[width=0.9\linewidth]{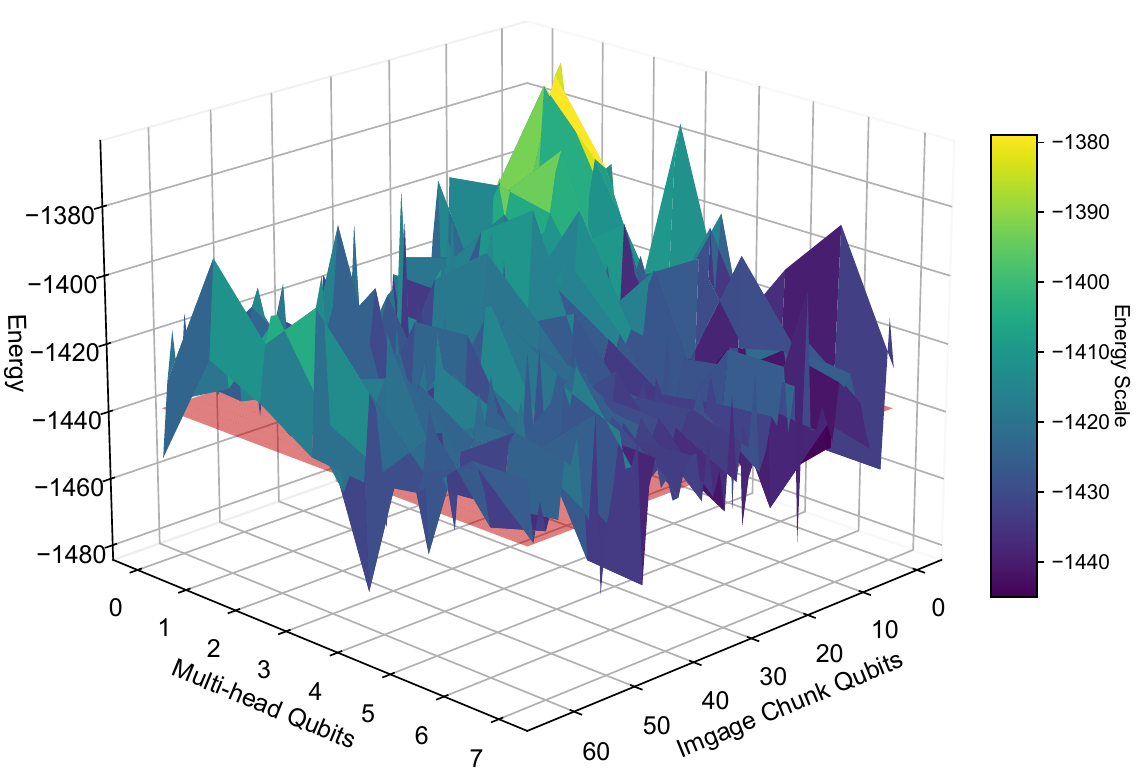}
    \caption{The state of the solution space after a single-bit mutation and the optimal solution plane of quantum annealing.}
    \label{fig:space}
\end{figure}

Figure~\ref{fig:space} visualizes the energy landscape resulting from a single-qubit mutation analysis. 
The 3D surface shows how small perturbations in qubit states affect the energy levels of the system. 
The red plane represents the current solution subspace with the full energy -1,438.95.
And the observable mutated energy surface, with a mean energy of -1,425.33, suggests that QAMA successfully navigates the energy landscape to reach near-optimal configurations. 
This validates the theoretical assumption that quantum annealing can effectively converge toward global or near-global minima in the high-dimensional attention space.

\begin{figure}[htbp]
    \centering
    \includegraphics[width=\columnwidth]{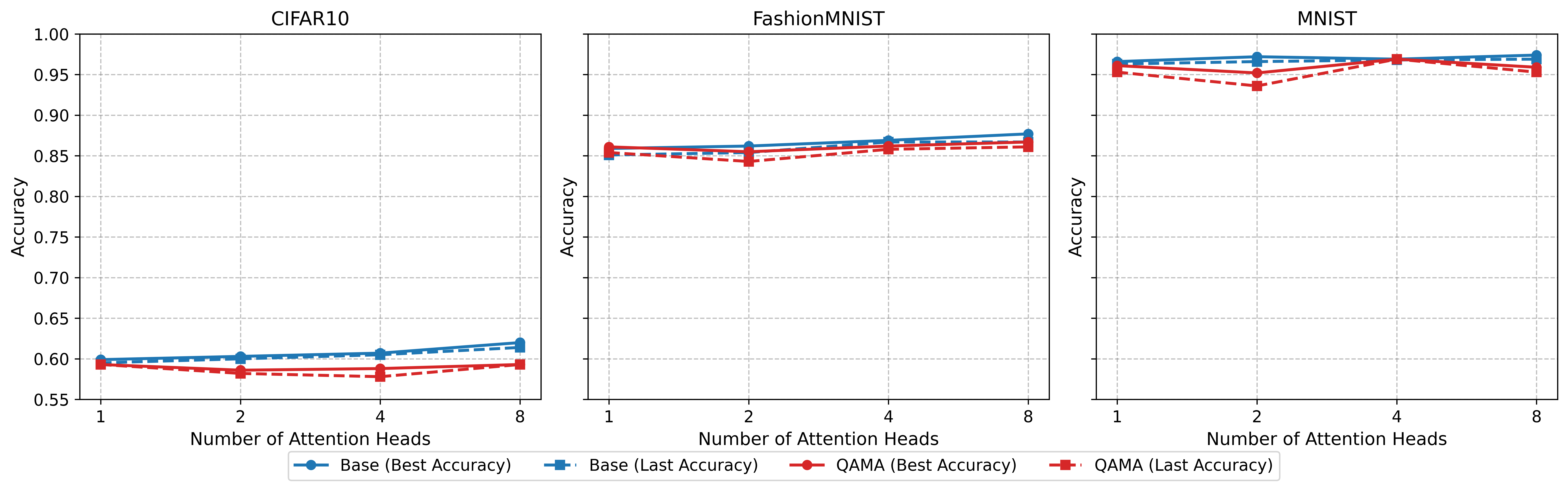}
    \caption{Model generalization experiments validating the accuracy performance of QAMA versus the original baseline model under different multi-head configurations.}
    \label{fig:heads}
\end{figure}

\begin{table}[htbp]
\centering
\caption{Performance differences due to the number of heads}
\label{tab:heads}
\resizebox{\columnwidth}{!}{%
\begin{tabular}{cccccc}
\hline
\multirow{2}{*}{Datasets} &
  \multirow{2}{*}{Heads} &
  \multirow{2}{*}{Qubits} &
  \multicolumn{2}{c}{Best Accuracy} &
  \multirow{2}{*}{Delta} \\ \cline{4-5}
                         &   &     & Base  & QAMA  &        \\ \hline
\multirow{4}{*}{MNIST}   & 1 & 49  & 0.966 & 0.961 & 0.51\% \\
                         & 2 & 98  & 0.972 & 0.952 & 2.05\% \\
                         & 4 & 196 & 0.969 & 0.969 & 0.00\% \\
                         & 8 & 392 & 0.974 & 0.959 & 1.54\% \\ \hline
\multirow{4}{*}{CIFAR10} & 1 & 64  & 0.599 & 0.593 & 1.00\% \\
                         & 2 & 128 & 0.603 & 0.586 & 2.81\% \\
                         & 4 & 256 & 0.607 & 0.588 & 3.13\% \\
                         & 8 & 512 & 0.62  & 0.593 & 4.35\% \\ \hline
\multirow{4}{*}{\begin{tabular}[c]{@{}c@{}}Fashion\\ MNIST\end{tabular}} &
  1 &
  49 &
  0.859 &
  0.861 &
  0.23\% \\
                         & 2 & 98  & 0.862 & 0.855 & 0.81\% \\
                         & 4 & 196 & 0.869 & 0.862 & 0.80\% \\
                         & 8 & 392 & 0.877 & 0.867 & 1.14\% \\ \hline
\end{tabular}%
}
\end{table}

\begin{figure}[htbp]
    \centering
    \includegraphics[width=\columnwidth]{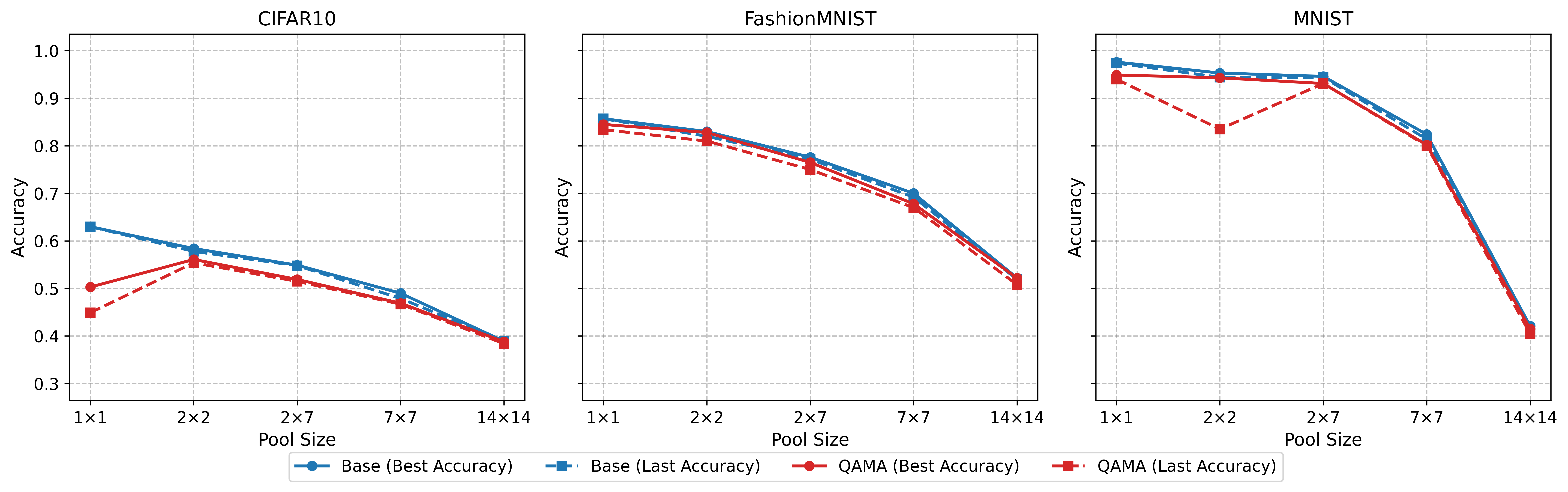}
    \caption{Model generalization experiments demonstrating the accuracy of QAMA versus the original model by testing various data lengths achieved through the use of different pooling sizes.}
    \label{fig:pools}
\end{figure}

\begin{table}[htbp]
\centering
\caption{Performance differences due to the length}
\label{tab:pools}
\resizebox{\columnwidth}{!}{%
\begin{tabular}{ccccccc}
\hline
\multirow{2}{*}{Datasets} &
  \multirow{2}{*}{Pools} &
  \multirow{2}{*}{Length} &
  \multirow{2}{*}{Qubits} &
  \multicolumn{2}{c}{Best Accuracy} &
  \multirow{2}{*}{Delta} \\ \cline{5-6}
                         &              &     &     & Base  & QAMA  &         \\ \hline
\multirow{5}{*}{MNIST}   & 1$\times$1   & 196 & 392 & 0.976 & 0.949 & 2.76\%  \\
                         & 2$\times$2   & 49  & 98  & 0.953 & 0.943 & 1.04\%  \\
                         & 2$\times$7   & 14  & 28  & 0.946 & 0.931 & 1.58\%  \\
                         & 7$\times$7    & 4   & 8   & 0.824 & 0.802 & 2.66\%  \\
                         & 14$\times$14 & 1   & 2   & 0.421 & 0.415 & 1.42\%  \\ \hline
\multirow{5}{*}{CIFAR10} & 1$\times$1   & 256 & 512 & 0.63  & 0.503 & 20.15\% \\
                         & 2$\times$2   & 64  & 128 & 0.584 & 0.561 & 3.93\%  \\
                         & 2$\times$7   & 16  & 32  & 0.549 & 0.519 & 5.46\%  \\
                         & 7$\times$7    & 4   & 8   & 0.49  & 0.469 & 4.28\%  \\
                         & 14$\times$14 & 1   & 2   & 0.389 & 0.388 & 0.25\%  \\ \hline
\multirow{5}{*}{\begin{tabular}[c]{@{}c@{}}Fashion\\ MNIST\end{tabular}} &
  1$\times$1 &
  196 &
  392 &
  0.857 &
  0.845 &
  1.40\% \\
                         & 2$\times$2   & 49  & 98  & 0.83  & 0.828 & 0.24\%  \\
                         & 2$\times$7   & 14  & 28  & 0.776 & 0.765 & 1.41\%  \\
                         & 7$\times$7    & 4   & 8   & 0.7   & 0.678 & 3.14\%  \\
                         & 14$\times$14 & 1   & 2   & 0.522 & 0.522 & 0.00\%  \\ \hline
\end{tabular}%
}
\end{table}

To conduct a detailed study of QAMA’s scalability across different input sizes and to establish performance consistency with the baseline model, we focused our investigation on the effects of head number and token sequence length.
Figure~\ref{fig:heads} presents the results for configurations with 1, 2, 4, and 8 attention heads. 
Performance is observed to increase incrementally with a higher number of heads, stabilizing at a level consistent with the performance of the classical baseline. 
Furthermore, Figure~\ref{fig:pools} illustrates the changes in performance when the input dimension is altered via different pooling kernel sizes.
Specifically, the increased kernel size results in a reduced input token length and a corresponding rapid degradation in model performance.
Crucially, this degradation mirrors the performance trajectory observed in the classical benchmark. 
According to Table~\ref{tab:heads} and Table~\ref{tab:pools}, it can find that the average accuracy differences are 3.32\% and 1.53\% on average, which is enough to show the performance consistency of QAMA and the classic benchmark.

In summary, the integration of these visual analyses provides concrete evidence of QAMA’s functional soundness, optimization efficiency, and interpretability in both NLP and CV domains. 
The sparse and targeted attention, energy-mapping feature transmission, and effective energy minimization collectively highlight the advantages of leveraging quantum annealing in attention-based architectures.

\subsection{Real machine experiment of CIM quantum annealer}\label{sec:cim_time}

\begin{table}[htbp]
\centering
\caption{CIM real machine experiment, compared to simulation.}
\label{tab:CIM}
\resizebox{\columnwidth}{!}{
\begin{tabular}{@{}llll@{}}
\toprule
                     & CIM    & Simulate & Delta           \\ \midrule
Average runtime      & 162 us & 402 ms   & $\approx$401 ms \\
Accuracy             & 97.5\% & 97.5\%   & 0.0\%           \\
Average label logits & 0.9664 & 0.9647   & 0.0017          \\ \bottomrule
\end{tabular}
}
\end{table}

The experimental outcomes of Table~\ref{tab:CIM} confirm that the QAMA operator not only functions effectively on quantum hardware but also demonstrates a significant practical advantage, particularly in computational speed.

The most striking result is the reduction in inference time. 
The average runtime on the CIM hardware was approximately 162 microseconds (µs), whereas the classical simulation required an average of 402 milliseconds (ms) per sample. 
This represents a massive speedup, with the quantum annealer completing the computation about 2,481 times faster than the simulation.
This substantial acceleration highlights the key strength of executing the QAMA operator on its native hardware, confirming the theoretical benefits of its design in a real-world application.

Crucially, the significant speedup was achieved with no loss in performance accuracy.
The model achieved an accuracy of 97.5\% on both the CIM device and in the simulation, resulting in a delta of 0.0\%.
This demonstrates that the QAMA operator's mapping to the CIM architecture is robust and that the hardware can solve the underlying optimization problem with high fidelity.
Further supporting this, the average label logits, which directly reflect the confidence level of model, exhibited near-perfect alignment.
The CIM produced an average logit of 0.9664, compared to 0.9647 in the simulation—a negligible difference of just 0.0017.
This confirms that the output quality and model confidence are preserved when transitioning from simulation to real quantum hardware.

In summary, the deployment on the QBoson CPQC-550 CIM annealer validates the QAMA operator's practical applicability.
It delivers on its theoretical promise, providing a substantial computational speedup while maintaining the high accuracy established during simulated training. 
This successful transition from simulation to hardware marks a critical step in demonstrating the real-world advantages of quantum-enhanced attention mechanisms.

\subsection{QAMA outperforms classic improved attention in complexity}

\begin{table*}[htbp]
\centering
\caption{Improved attention mechanism comparison.}
\label{tab:improved_attention}
\resizebox{\textwidth}{!}{
\begin{tabular}{cccccc}
\hline
\multirow{2}{*}{Category}         & \multirow{2}{*}{Method} & \multicolumn{2}{c}{Complexity} & \multicolumn{2}{c}{Accuracy} \\ \cline{3-6} 
                        &           & Time             & Space         & Last   & Best   \\ \hline
Baseline                & Attention & $O(n^2)$         & $O(n^2)$      & 61.4\% & 62.0\% \\
Quantum annealing       & QAMA      & $T_{sol}$ & $O(n)$        & 59.3\% & 59.3\% \\
\multirow{2}{*}{Sparse attention} & Longformer              & $O(nw+ng)$     & $O(nw+ng)$    & 62.1\%        & 62.3\%       \\
                        & BigBird   & $O(n(g+w+r))$    & $O(n(g+w+r))$ & 60.0\% & 61.0\% \\
Low-rank approximations & Linformer & $O(nk)$          & $O(nk)$       & 59.2\% & 59.4\% \\
Kernelized attention    & Performer & $O(Lmd)$         & $O(md+Ld+mL)$ & 59.7\% & 60.0\% \\
Hashing-based attention & Reformer  & $O(n^2)$         & $O(n)$        & 62.5\% & 62.9\% \\ \hline
\end{tabular}
}
\end{table*}

When compared to other attention mechanisms optimized for spatiotemporal complexity, the QAMA operator's primary advantages are its unique theoretical time complexity, derived from quantum annealing principles, and its highly efficient space complexity, while maintaining competitive performance accuracy.
As shown in Table~\ref{tab:improved_attention}, QAMA introduces a fundamentally different complexity profile compared to classical approaches.
Space Complexity: QAMA achieves a space complexity of $O(n)$, matching the highly optimized Reformer. 
This is a significant improvement over the baseline's $O(n^2)$ and is more efficient than improved methods, whose space requirements scale with window or block sizes.
Time Complexity: QAMA's time complexity of $T_{sol}$ is unique. 
It does not depend on the sequence length n but rather on the problem's Hamiltonian. 
This offers a potential paradigm shift, suggesting that for certain problem structures, QAMA could theoretically offer constant-time attention, a feat unattainable by classical methods that inherently scale with sequence length in some form.

The results on CIFAR10 demonstrate that QAMA's theoretical efficiency does not sacrifice a substantial amount of accuracy.
With a peak accuracy of 59.3\%, QAMA delivered highly competitive performance against advanced methods like Linformer (59.4\%) and Performer (60.0\%). 
Although methods like Longformer (62.3\%) and Reformer (62.9\%) achieved higher accuracy in this specific configuration, QAMA's performance is noteworthy. 
It successfully avoids the severe performance degradation that can sometimes accompany complexity optimizations, positioning it as a robust and reliable operator.

In conclusion, the primary advantage of QAMA lies in its holistic value proposition. 
It combines a revolutionary theoretical time complexity, superior space efficiency, and a demonstrated ability to maintain competitive accuracy. 
This makes it a compelling alternative to classical optimized attention mechanisms, offering not just an incremental improvement but a new pathway for building highly efficient and scalable models.

\section{Conclusion}\label{sec:conclusion}

This paper proposes QAMA, a novel multi-head attention operator that utilizes quantum annealing to tackle the efficiency limitations found in conventional attention mechanisms.
The core innovation of QAMA lies in reformulating the attention mechanism as a combinatorial optimization problem. 
By then constructing an energy-based Hamiltonian system based on this formulation, QAMA is able to utilize quantum annealers for highly efficient attention computation.
Our approach achieves a time complexity of $T_{sol}$ and a space complexity of $O(n)$.
Our analysis using TTS proves that this represents a fundamentally different algorithm and implementation compared to classical time complexity, realizing an efficient attention mechanism on CIM hardware.
Extensive experiments across NLP and CV tasks demonstrate that QAMA maintains performance parity with classical attention mechanisms while providing significant theoretical and practical advantages. 
Furthermore, successful deployment on a real quantum annealing machine validates the feasibility and efficiency of QAMA in real-world scenarios, showcasing notable computational speedups and practicality.
QAMA represents a crucial development in the integration of quantum computing and deep learning, laying the groundwork for more scalable and efficient AI models.
With the ongoing evolution of quantum computation, QAMA demonstrates the potential of quantum-enhanced architectures to address the constraints of classical computing paradigms.

\section*{Acknowledgments}
This work was supported in part by the National Natural Science Foundation of China under Grant 62272483;
in part by the Natural Science Foundation for Distinguished Young Scholars of Hunan Province under Grant 2023JJ10078;
in part by the CCF-QBoson Quantum Computing Application Innovation Fund (NO.CCF-Boson202404);
in part by the Monumental Consultation Project on the Development Strategy of Chinese Engineering and Technology (2025WK1001).



 
%

\bibliographystyle{IEEEtran}
\bibliography{refs}


\section{Biography Section}
\vspace{-1cm}

\begin{IEEEbiographynophoto}{Peng Du}
received his B.S. degree in College of Information, Huazhong Agricultural University, Wuhan, China, in 2024. He is pursuing his M.S. degree at the School of Electronic Information, Central South University, Changsha, China.
\end{IEEEbiographynophoto}
\vspace{-1cm}

\begin{IEEEbiographynophoto}
{Professor Jinjing Shi}(Senior Member, IEEE) is now a professor in the School of Electronic Information of Central South University. She received her B.S. and Ph.D. degrees in the School of Information Science and Engineering, Central South University, Changsha, China, in 2008 and 2013, respectively. She was selected in the ”Shenghua lieying” talent program of Central South University and Special Foundation for Distinguished Young Scientists of Changsha in 2013 and 2019, respectively. Her research interests include quantum computation and quantum cryptography. She has presided over the National Natural Science Foundation Project of China and that of Hunan Province. There are 50 academic papers published in important international academic journals and conferences. She has received the second prize of natural science and the outstanding doctoral dissertation of Hunan Province in 2015, and she has received the Best Paper Award in the international academic conference MSPT2011 and Outstanding Paper Award in IEEE ICACT2012.
\end{IEEEbiographynophoto}

\vspace{-1cm}

\begin{IEEEbiographynophoto}
{Wenxuan Wang}(Student Member, IEEE) received her B.S. degree in College of Physical Science and Technology, Central China Normal University, Wuhan, China, in 2020 and M.S. degree in the School of Computer Science and Engineering, Central South University, Changsha, China. She is currently a doctor student at the School of Computer Science and Engineering, Central South University, Changsha, China. Her research interests mainly include Hamiltonian learning, quantum state preparation, and quantum machine learning.
\end{IEEEbiographynophoto}

\vspace{-1cm}

\begin{IEEEbiographynophoto}{Yin Ma}
is the Founder and COO of Beijing QBoson Quantum Technology Co. Ltd.
\end{IEEEbiographynophoto}

\vspace{-1cm}

\begin{IEEEbiographynophoto}{Kai Wen}
is the Founder and CEO of Beijing QBoson Quantum Technology Co. Ltd.
\end{IEEEbiographynophoto}

\vspace{-1cm}
	

\begin{IEEEbiographynophoto}{Professor Xuelong Li} (Fellow, IEEE) is the CTO and Chief Scientist of China Telecom, where he founded the Institute of Artificial Intelligence (TeleAI) of China Telecom.
\end{IEEEbiographynophoto}

\vfill

\end{document}